\documentclass[%
 aps,
 amsmath,amssymb,
 reprint,%
]{revtex4-1}

\usepackage{graphicx}
\usepackage{dcolumn}
\usepackage{bm}

\usepackage[utf8]{inputenc}
\usepackage[T1]{fontenc}
\usepackage{mathptmx}
\usepackage{etoolbox}
\usepackage{hyperref}
\usepackage{mhchem}
\usepackage{xcolor}

\makeatletter
\def\@email#1#2{%
 \endgroup
 \patchcmd{\titleblock@produce}
  {\frontmatter@RRAPformat}
  {\frontmatter@RRAPformat{\produce@RRAP{*#1\href{mailto:#2}{#2}}}\frontmatter@RRAPformat}
  {}{}
}%
\makeatother
\begin{document}

\preprint{AIP/123-QED}

\title[Large-area CVD Graphene for Photogating-Based Photodetection]{ Large-area CVD Graphene for Photogating-Based Photodetection}
\author{D.A. Matienko$^{1,*}$,  D.P. Borisenko$^{2}$,A.Yu. Kuntsevich$^{1}$},
\affiliation{ 
$^1$- P.N. Lebedev Physical Institute of the Russian Academy of Sciences. Leninsky Prospekt 53, 119991, Moscow, Russia\\
$^2$- National Research Nuclear University “MEPhI”. Kashirskoe highway 31, 115409, Moscow, Russia\\
}%
\email{d.matienko@lebedev.ru}

\date{\today}

\begin{abstract}

Our work explores large-area CVD graphene as a scalable platform for photodetectors. 
We show that substrate-induced photogating in graphene/SiO$_2$/Si structures enables a photoconductivity signal in mono-, bi-, and trilayer CVD graphene. 
The devices respond to illumination with photon energies above the Si band gap and do not respond to telecom frequency, suggesting the photoconductivity mechanism through the photogating. Large sample area and gate voltage-dependent resistivity measurements allow us to prove the dominant role of the photogating directly, demonstrating similarity of the capacitive recharging current and the photoresponse. 
The sensitivity is the highest \(995~\mathrm{A/W}\) for the monolayer graphene. Increasing the number of graphene layers reduces the sensitivity due to screening and parallel conduction, but can improve detectivity by lowering the noise level. 
The sensitivity of CVD graphene device depends on fabrication route of photolitography:  device with metal contacts fabricated before graphene transfer show higher photoresponce.
Thus, our work demonstrate the potential of industrially relevant CVD graphene for sensitive photogating-based photodetectors.

\end{abstract}

\maketitle

\section{Introduction}

Graphene is one of the basic 2D electronic materials. Chemical vapor deposition (CVD) technology enables the growth controllably almost uniform one- or few-layer polycrystalline graphene wafer-scale layers~\cite{li2009large,reina2009large}. 
Their mechanical transfer on arbitrary substrates opens the pathway to electronic applications ~\cite{gomezdearco2010continuous, wu2011high, defazio2016high}. 

 Gapless nature of the graphene monolayer spectrum leads to absorption of about 2\% of light in wide range of wavelength from Thz to UV due to interband optical transiotions ~\cite{nair2008fine,kuzmenko2008universal,mak2008measurement}. Graphene photodetectors have therefore become a promising direction in future optoelectronics, owing to their broadband operation, high carrier mobility, and compatibility with ultrafast and flexible devices~\cite{bonaccorso2010graphene,xia2014graphene}.
 
Several mechanisms can contribute to the photoresponse of graphene-based devices, including photoexcitation of carriers, photovoltaic, photothermoelectric effects, bolometric response and photogating~\cite{koppens2014photodetectors,freitag2013photoconductivity,guo2016high}.

Incident light excites carries in graphene and electron system exhibits fast (sub-ps) relaxation to the nonequilibrium thermalization with enhanced electron temperature,  while maintaining the lattice relatively cold. Therefore there is no strong bolometric effect in graphene. Some bolometric effect, linear in intensity, is possible due to residual lattice heating and under applied bias ~\cite{freitag2013photoconductivity,jago2019microscopic}.

In homogeneous graphene, direct photoexcitation alone usually gives a limited photoresponse because of the absence of a band gap and fast carrier relaxation. 

In metal--graphene--metal devices, the photo-EMF is generated near contacts. 
Built-in electric fields and contact-induced doping produce photovoltaic contributions, whereas local heating of the electronic system generates photothermoelectric voltage due to spatial variations of the Seebeck coefficient. 
These photoelectric and photothermoelectric mechanisms often coexist and are difficult to separate experimentally~\cite{echtermeyer2014photothermoelectric,bandurin2018dual}.

Within the photogating mechanism, light is absorbed not necessarily in graphene itself, but in a nearby material or at an interface coupled to graphene. 
The photoexcited charges are separated or trapped, producing a long-lived electrostatic perturbation that acts as an effective gate voltage. 
Graphene then serves as a highly conductive and gate-sensitive readout channel, whose conductance changes under this photoinduced gating. 
This mechanism can provide large photoconductive gain and high sensitivity~\cite{konstantatos2012hybrid, wang2021interfacial}.

Understanding the nature of the graphene photoresponse is often complicated by the coexistence of several mechanisms within the same sample. 
Most studies reported in the literature focus either on mechanically exfoliated graphene micrometer-scale devices or on modified CVD graphene structures. 
Meanwhile, the photoconductivity of technologically simple, unmodified CVD graphene on Si/SiO$_2$ substrates remains much less systematically explored. 
Our work aims to fill this gap. 

We demonstrate that the photoconductive response of this system is governed predominantly by Si-induced photogating rather than by direct light absorption in graphene. 
This conclusion is supported by the sublinear intensity dependence, the absence of a detectable response at 1550~nm, and the similarity between the optical-power dependences of the graphene photoresponse and the photoinduced recharging current in the graphene/SiO$_2$/Si capacitive structure. The silicon substrate acts as a photoactive underlayer and produces an effective back-gate action on graphene, enabling high sensitivity at low optical power density. 
At higher illumination levels, the response gradually saturates like a human eyes, providing a broad dynamic range that may be useful for practical photodetector operation.

\section{Methods}

\begin{figure*}[t]
    \centering
    \includegraphics[width=\textwidth]{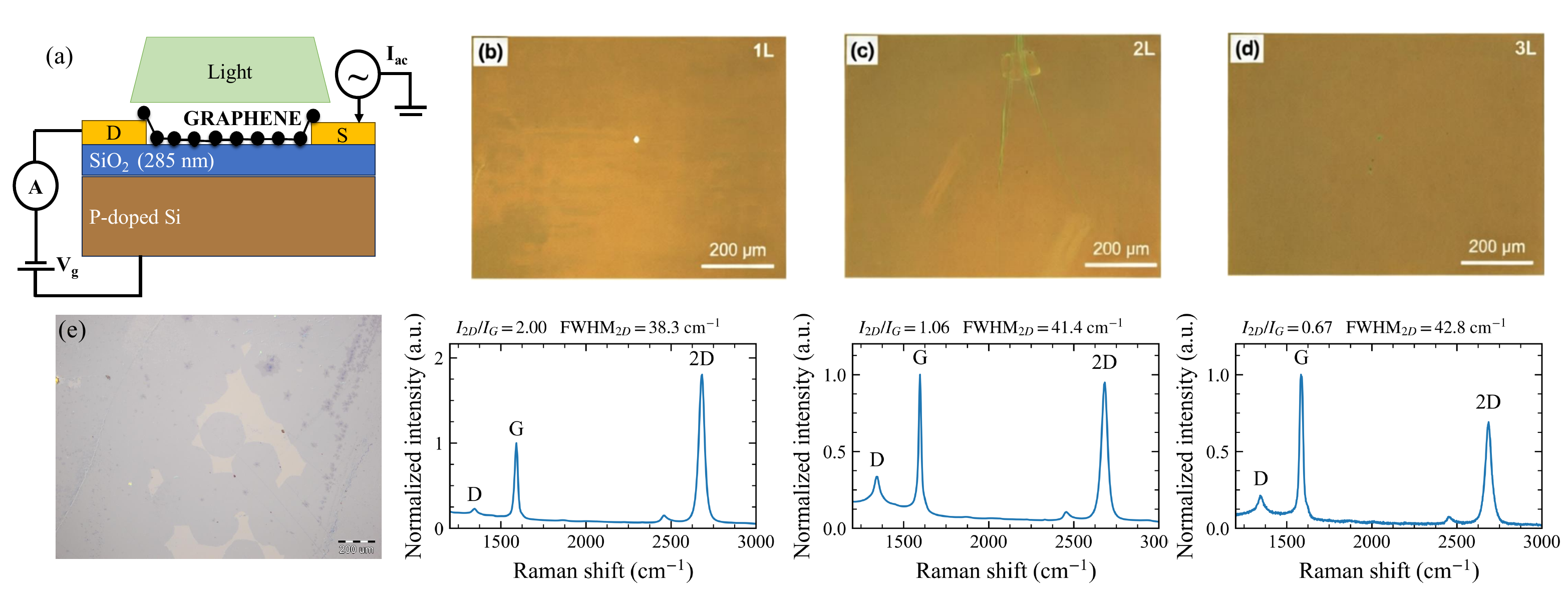}
    \caption{
    Structure and Raman characterization of the graphene devices.
    (a) Schematic of the graphene/SiO$_2$/Si device and top view of the active region.
    (b--d) Optical microscopy images top of the 1L, 2L, and 3L graphene samples, respectively. The Raman spectra for the corresponding parts are shown in the bottom panels.
    (e) Optical microscopy image of the active device region.
    }
    \label{fig:sample_raman}
\end{figure*}

\subsection{CVD growth of graphene}

Graphene films were grown by chemical vapor deposition (CVD) on the surface of electrochemically polished copper foil with a purity of 99.9\%, an RMS roughness below 5~nm, and a thickness of approximately \(25~\mu\mathrm{m}\).  The growth was carried out in a cold-wall CVD reactor developed by RusGraphene LLC ~\cite{rybin2024rapid,borisenko2025raman}.  The copper foil sample was fixed between two electrodes and heated by an electric current.  The temperature was monitored through an observation window using a pyrometer located outside the CVD reactor. 
To improve the surface quality of the copper catalyst and promote the subsequent growth of predominantly single-layer graphene, the catalyst was annealed in an \(\mathrm{Ar/H_2}\) atmosphere at temperatures above \(900^\circ\mathrm{C}\) for 30~min. 
After that, methane was introduced into the CVD reactor at a chamber pressure of 100 mbar, followed by a growth step of up to 5~min to form a continuous single-layer graphene film on the catalyst surface.

\subsection{Transfer of graphene}

The transfer of the obtained graphene films to Si/SiO$_2$ substrates was carried out using a polymer frame PMMA 950 A4 by wet transfer. A polymer frame layer was applied to the surface of a copper catalyst with a graphene film, followed by the removal of the graphene film from the back of the catalyst in an oxygen plasma. The removal of the copper catalyst was carried out using a 10 \% aqueous solution of ammonium persulfate for 1 hour. After completing the etching step of the catalyst, PMMA/graphene was moved to a separate container with deionized water for 15 minutes, followed by transfer to the surface of the target substrate Si/SiO$_2$. The PMMA layer was removed from the graphene surface using acetone and isopropanol. As a result, a large-area CVD graphene monolayer was obtained on the target substrate. 95\% coverage area was graphene monolayer (crystallites with lateral dimensions about 200 $\mu$m) and the remaining 5\% was either empty or bilayer. The typical microscope image of the graphene film is shown in Fig.~\ref{fig:sample_raman}(e) 

Multilayer graphene structures were obtained by repeating the transfer procedure. As a result, step-like structures containing one-, two-, and three-layer graphene regions were formed on the same substrate. Optical images of the fabricated one-, two-, and three-layer graphene structures are shown in Fig.~\ref{fig:sample_raman}(b--d).

\subsection{Device fabrication}

  Cr/Au source and drain contacts were formed by mask photolithography, vacuum evaporation, and lift-off. Resist mask consisting of LOR lift-off resist and FP2506 photoresist was used to define the contact pattern. The metallization consisted of a 5~nm Cr adhesion layer and a 65~nm Au conducting layer.

Two fabrication routes were used in this work with different in the order of graphene transfer and contact fabrication. In the first route, CVD graphene was transferred onto the Si/SiO$_2$ substrate before contact fabrication and Cr/Au contacts were deposited directly on the graphene layer. In the second route, the Cr/Au contacts were first fabricated on the bare Si/SiO$_2$ substrate and then CVD graphene was transferred onto the pre-patterned metal contacts. 

The graphene layer number was confirmed by Raman spectroscopy, as shown in Fig.~\ref{fig:sample_raman}. 
The spectra show the characteristic D, G, and 2D bands. 
The decrease in \(I_{2D}/I_G\) from \(2.00\) for 1L graphene to \(1.06\) for 2L and \(0.67\) for 3L, together with the broadening of the 2D band from \(38.3\) to \(42.8~\mathrm{cm^{-1}}\), confirms the transition from monolayer to trilayer graphene.

\subsection{Sample geometry and mounting}

A schematic view of the device geometry is shown in Fig.~\ref{fig:sample_raman}(a). The graphene channel was located between the source and drain electrodes. The size of the active graphene region was $1 \times 1$~mm$^2$. During measurements, light spot covered the whole graphene channel between the electrodes.

After fabrication, the substrates were cut into individual samples and bonded for measurements. Electrical connection to the contact pads was made using gold 20 $\mu$m diameter wires fixed with silver paste. We glued a contact to the silicon substrate used as a back gate. 

\subsection{Photoconductivity measurements} 

Photoconductivity measurements were performed using a dual-frequency modulation scheme using two SR830 lock-in amplifiers. 

The first lock-in amplifier applied AC electrical excitation between the source and drain electrodes and detected the electrical conductivity response of the graphene channel at \(177~\mathrm{Hz}\). We used about 0.5 mA driving current with the typical resistance being about several hundreed Ohms.
 The output signal of the first lock-in amplifier was then analyzed by the second lock-in amplifier referenced to the chopper frequency of \(16~\mathrm{Hz}\), used to modulate CW laser excitation.
The signal from the second lock-in amplifier gives directly the photoconductance of the device.

The incident optical power was adjusted by mutual rotation of two polarizers and measured using a two calibrated power meter (one for nM-to-$\mu$W range and another for $\mu$W-to mW range). We used a set of diode and DPSS lasers with a power below 100 mW (405 nm, 532 nm, 635 nm, 808 nm, 1550 nm).

For gate-dependent measurements, the silicon substrate was used as a global back gate. 
Keithley 6517A electrometer applied the gate voltage \(V_g\) to the silicon substrate 
and monitored the leakage current of the sample.

All measurements were carried out at room temperature on air.


\section{Results}

In total, 15 graphene devices were fabricated and investigated in this work. 
The photoresponse trends discussed below were reproducible across different samples. 
The data in the paper are representative measurements selected for clarity.
\subsection{Effect of top/bottom contacts on the photoresponse}

\begin{figure}[ht]
    \centering
    \includegraphics[width=1\linewidth]{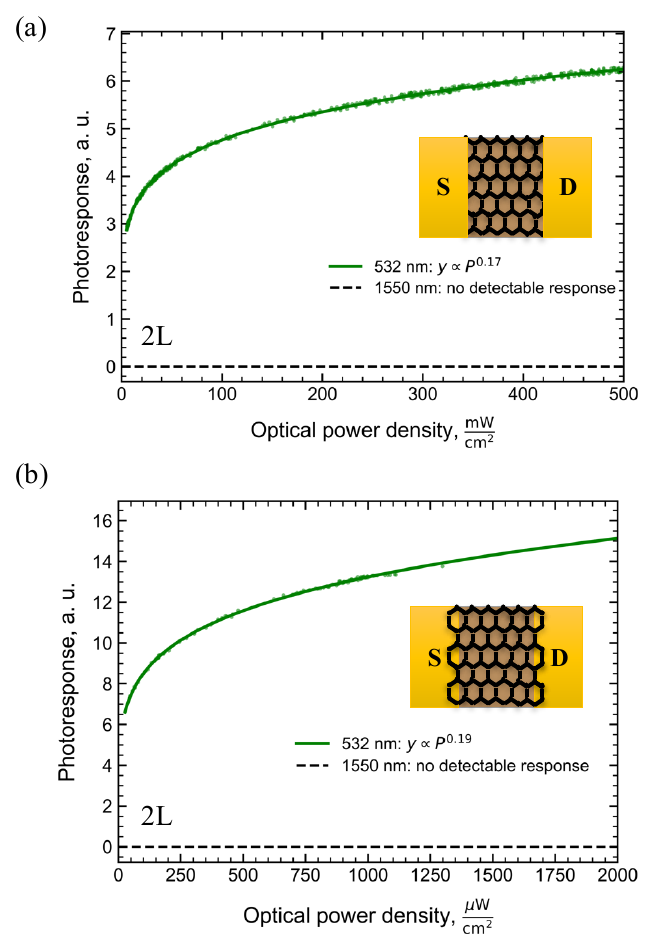}
    \caption{Dependence of the photoresponse on optical power density for 2 layer CVD graphene fabricated using two different fabrication routes: (a) metal contacts deposited on top of graphene and (b) graphene transferred onto pre-patterned metal contacts.}
    \label{fig:PH_PL_contacts}
\end{figure}

 We compared two fabrication routes: (i) metal contacts deposited on top of graphene and (ii) graphene transferred onto pre-patterned metal contacts. The representative dependences of the photoresponse at 532 nm and 1550 nm on optical power density are shown in Fig.~\ref{fig:PH_PL_contacts}.

In both structures, the photoresponse $y$ at 532 nm increases with optical power density $P$ and follows a strongly sublinear dependence. The data could be well approximated by a power law $y = P^{\alpha}$.
In both cases,  \(\alpha \ll 1\) (\(\alpha = 0.18 \pm 0.01\)), indicating strong saturation of the photoresponse with increasing optical power density which is driven by the same physical mechanism.

The main difference between the data for two samples in Fig.~\ref{fig:PH_PL_contacts} is the horizontal scale. 
In the device with metal contacts deposited on top of graphene, saturation is observed in the mW/cm$^2$ range. In contrast, in the device where graphene was transferred onto pre-patterned contacts, saturation starts already in the \(\mu\)W/cm$^2$ range.

This enhancement is clearly due to a cleaner active graphene channel, since in bottom contact case graphene is not directly exposed to the contact lithography and lift-off processes. Indeed polymer and photoresist residues are well known to modify the doping level of CVD graphene, increase hysteresis, and reduce carrier mobility~\cite{Suk2013polymer,Suhail2017residue,Tyagi2022ultraclean}. 
The transfer of graphene onto pre-patterned substrates has also been demonstrated as a fabrication approach that limits the direct exposure of graphene to contact lithography~\cite{Abhilash2015transfer,cha2022bottom,ievleva2025metastable}.

For both device geometries, no detectable photoresponse was observed under 1550 nm illumination, as shown by the dashed curves in Fig.~\ref{fig:PH_PL_contacts}. 
The photon energy at this wavelength is lower than the band gap of silicon; therefore, 1550 nm light does not efficiently generate carriers in the Si substrate. 
This result shows that the photoresponse observed at 532 nm cannot be explained by direct absorption in graphene alone. 
Instead, it indicates that the Si/SiO$_2$ substrate plays an active role in the formation of the signal, which is consistent with a substrate-induced photogating mechanism.

Due to much higher photoresponse we discuss below the results for graphene transferred onto pre-patterned metal contacts. 

\subsection{Spectral dependence of the photoresponse}

\begin{figure}[ht]
    \centering
    \includegraphics[width=1\linewidth]{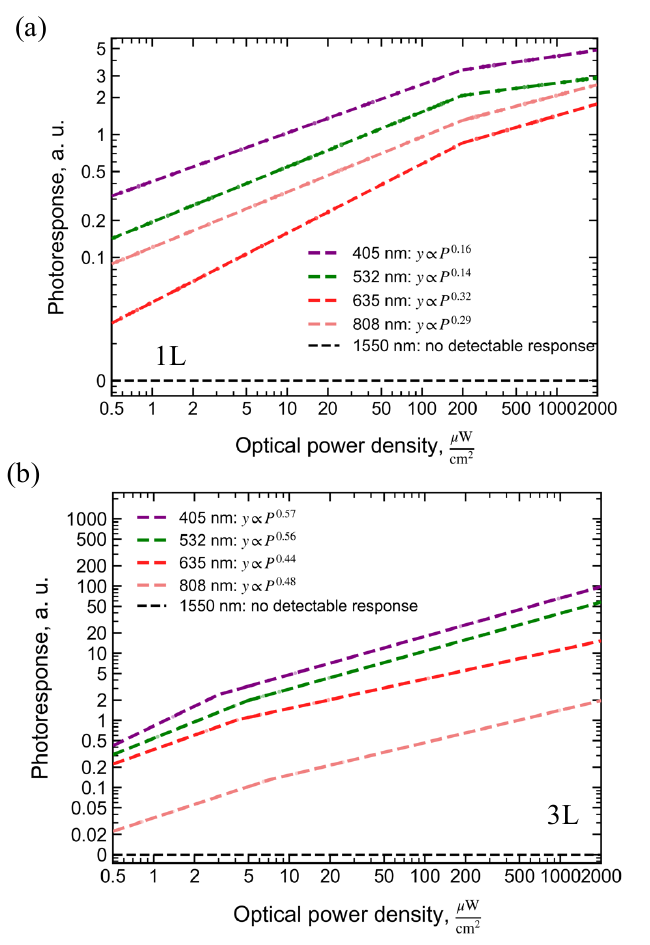}
    \caption{Dependence of the photoresponse on optical power density for monolayer (top panel) and trilayer (bottom panel) CVD graphene at different wavelengths.Dashed lines for high power density show power-law fits, \(y = AP^{\alpha}\), with the corresponding exponents given in the legend. Dashed lines for low power density indicate the linear low-power regions used to extract the sensitivity and detectivity.}
    \label{fig:PH_PL_spectral}
\end{figure}
 
Figures~\ref{fig:PH_PL_spectral}a,b show the dependencies of the photoresponse on optical power density for various excitation wavelengths for representative mono- and trilayer samples, respectively.

For both devices, a measurable photoresponse is observed at 405, 532, 635, and 808~nm, whereas illumination at 1550~nm does not produce a detectable signal. 
The absence of the response at 1550~nm is an important feature, since the photon energy at this wavelength is below the band gap of silicon. 
Therefore, the observed photoresponse at shorter wavelengths is not determined by direct absorption in graphene alone and is strongly affected by the Si/SiO$_2$ substrate.

For all wavelengths with a measurable signal, the photoresponse increases with optical power density and follows a sublinear power-law dependence. 
For monolayer graphene, the extracted exponents are  \(0.14{-}0.32\). 
For trilayer graphene, the corresponding exponents are higher and lie in the range \(0.48{-}0.57\). 
Thus, the saturation is more pronounced in the monolayer device, while the trilayer device shows a weaker nonlinearity.

At low optical power densities, the initial parts of the photoresponse curves are well approximated by a linear dependence as shown by dashed lines in Fig.~\ref{fig:PH_PL_spectral}. 
They were used to determine the spectral sensitivity \(S\) of the devices according to
$S= I_{\mathrm{ph}}/P_{\mathrm{opt}}$, where \(I_{\mathrm{ph}}\) is the light-induced current, \(P_{\mathrm{opt}}\) is the optical power incident on the device. The specific detectivity was calculated as $D^*=S\sqrt{A_{\mathrm{act}}\Delta f}/i_{\mathrm{n}}$, where \(A_{\mathrm{act}}\) is the illuminated active area, \(i_{\mathrm{n}}\) is the dark current noise measured without illumination, and \(\Delta f\) is the effective lock-in detection bandwidth.

The extracted sensitivity and detectivity values are summarized in Table~\ref{tab:spectral_sensitivity_detectivity}. 
The monolayer device demonstrates higher sensitivity at all investigated wavelengths. 
The maximum value, \(S = 995~\mathrm{A/W}\), is obtained at 635~nm. 
For trilayer graphene, the sensitivity is lower and ranges from \(22.8\) to \(73.4~\mathrm{A/W}\). 

The decrease in sensitivity with number of graphene layers is due to parallel conection of several graphene layers and partial screening of the photoinduced electric field from the substrate by the first layer. 

At the same time, the detectivity does not drop strongly with number of layers. 
In particular, at 808~nm the detectivity of the trilayer is even higher than the one for the monolayer device. This is because the lower noise level in trilayer structure compensates for the reduce in the sensitivity.

\begin{table}
\caption{\label{tab:spectral_sensitivity_detectivity}
Spectral sensitivity \(S\) and detectivity \(D^*\) of mono- and trilayer CVD graphene photodetectors.}
\begin{ruledtabular}
\begin{tabular}{ccccc}
$\lambda$ 
& $S_{\mathrm{1L}}$ 
& $D^*_{\mathrm{1L}}$ 
& $S_{\mathrm{3L}}$ 
& $D^*_{\mathrm{3L}}$ \\
$(\mathrm{nm})$ 
& $\left(\frac{\mathrm{A}}{\mathrm{W}}\right)$ 
& $\left(10^{8}\,\mathrm{Jones}\right)$ 
& $\left(\frac{\mathrm{A}}{\mathrm{W}}\right)$ 
& $\left(10^{8}\,\mathrm{Jones}\right)$ \\
\hline
405 & 745 & 2.9 & 41.7 & 1.3 \\
532 & 752 & 2.9 & 22.8 & 0.7 \\
635 & 995 & 3.8 & 68.3 & 2.2 \\
808 & 433 & 1.7 & 73.4 & 2.4 \\
\end{tabular}
\end{ruledtabular}
\end{table}

\subsection{Photogating origin of the photoresponse}
 
The absence of a detectable photoresponse at 1550~nm in the spectral measurements points to a photogating mechanism.

In the studied graphene/SiO$_2$/Si devices, photogating can originate from both the graphene surface and the Si substrate. 
Surface adsorbates and polymer residues remaining after transfer and lithography could trap photoexcited charges and locally modulate the graphene resistance. 
Similar charge-trapping mechanisms are demonstrated in hybrid graphene photodetectors based on quantum dots, semiconductors, dielectric layers, or engineered interfaces~\cite{konstantatos2012hybrid,liu2014graphene,zhang2014ultrahigh,wang2021interfacial}. 

At the same time, photons with the energy above the bandgap generate electron--hole pairs in silicon, leading to a surface photovoltage~\cite{Kronik1999surface,Novikov2010experimental,guo2016high}. 
Photovoltage in Si/SiO$_2$ structures is a nonequilibrium phenomenon studied broadly in the past. It depends on many parameters, such as doping, compenstion, wavelength, temperature, SiO$_2$ thickness ~\cite{brattain1956combined,johnson1957measurement,quillet1960surface,goodman1961method,frankl1967theory,lam1971surfaceI,lam1971surfaceII}.

Note, the photon itself has no charge, and the photovoltage comes from the change in the thermodynamical workfunction (i.e. chemical potential) caused by nonequilibrium conditions. 
The gate voltage $V_g$ in graphene/SiO$_2$/Si structure is determined by the electrochemical potential difference:

\begin{equation}
    V_g =
    \left(\varphi_{\mathrm{Gr}}-\varphi_{\mathrm{Gate}}\right)
    +
    \frac{\mu_{\mathrm{Gr}}-\mu_{\mathrm{Gate}}}{e},
    \label{eq:electrochemical_gate_voltage}
\end{equation}

where \(\varphi\) are the electrostatic potentials, \(\mu\) are the chemical potential of graphene (Gr) and silicon (Si), respectively, and \(e\) is the elementary charge. 
The photoinduced potential is capacitively coupled to graphene through the oxide layer and therefore acts as an additional gate voltage.

\begin{figure}[ht]
    \centering
    \includegraphics[width=1\linewidth]{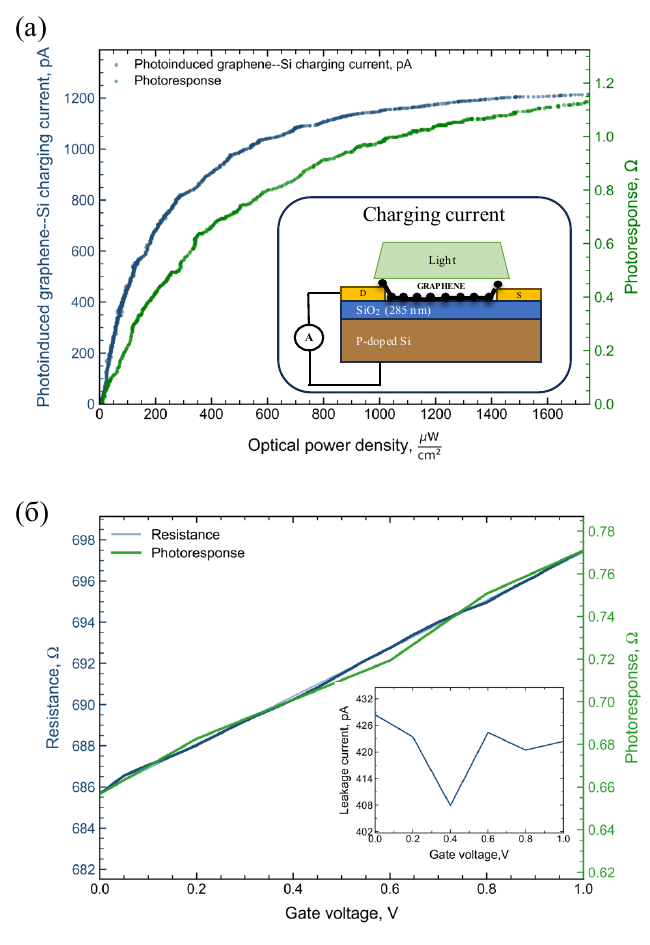}
    \caption{
Origin of the photoconductive response in the graphene/SiO$_2$/Si structure.
(a) Optical-power dependence of the graphene photoresponse and the graphene--Si charging current at 532~nm. The inset shows the charging-current measurement configuration.
(b) Gate-voltage dependence of the graphene resistance and photoresponse. The inset shows the leakage current throught the structure of SiO$_2$.
}
    \label{fig:photogating_origin}
\end{figure}

Two photogating mechanisms coexist; large sample area of CVD-graphene allows to perform an experiment crucious and identify the Si contribution. 
We use the the photovoltage measured through the recharging current between graphene and the Si substrate and the gate-voltage dependence of the graphene resistance. 
Indeed, for small photovoltages (below 0.3 V in our case), the graphene resistance change can be written as

\begin{equation}
    \Delta R_{\mathrm{ph}} =
    \frac{dR}{dV_g}\Delta V_{\mathrm{ph}},
    \label{eq:photogating_basic}
\end{equation}

where \(dR/dV_g\) is the gate sensitivity of the sample impedance and \(\Delta V_{\mathrm{ph}}\equiv (\mu_{Si Light}-\mu_{Si No Light})/e\) is the photovoltage. The photovoltage is related to recharging current (see scheme in the inset to Fig. ~\ref{fig:photogating_origin}a) as $\Delta V_{\mathrm{ph}} = {I_{\mathrm{ch}}}/(2\pi f C)$, where \(f\) is the light modulation frequency and \(C\) is the graphene-to-Si capacitance. A similar method was applied to study equilibrium thermodynamics through the derivatives of the chemical potential in Refs. \cite{teneh2012, kuntsevich2015}. 

The similar power dependence of graphene photoresponse and the photoinduced charging current in Fig.~\ref{fig:photogating_origin} suggests that the photoresistance of graphene reflects majorly photovoltage in Si. 
Using \(C_{\mathrm{ox}}=\varepsilon_0\varepsilon_{\mathrm{SiO_2}}S/d\), we obtain \(C_{\mathrm{ox}}\approx95\,\mathrm{pF}\). 
For optical power density 1.75 mW/cm$^2$ one has \(I_{\mathrm{ch}}\approx1.2\,\mathrm{nA}\) and \(f=16\,\mathrm{Hz}\), and, correspondingly \(\Delta V_{\mathrm{ph}}\approx0.125\,\mathrm{V}\). 
From Fig.~\ref{fig:photogating_origin}(b) one finds  \(dR/dV_g\approx11.7\,\Omega/\mathrm{V}\). 
Using Eq.~\eqref{eq:photogating_basic}, this gives \(\Delta R_{\mathrm{calc}}\approx1.5\,\Omega\) close to the measured saturation photoresponse, \(\Delta R_{\mathrm{exp}}\approx1.1\,\Omega\).

Thus, Si photogating is the main photoconductivity mechanism. The discrepancy between the calculated and measured photoresponse are related to the  uncertainty in capacitance  and graphene surface photogating.

\section{Discussion}

The photogating mechanism allows one to obtain a large electrical response despite the low effect of optical absorption on the graphene conductivity itself. 
The present work shows that this mechanism provides high sensitivity not only in exfoliated or specially engineered graphene structures, but also in large-area CVD graphene. 
This is important from a technological point of view, since CVD graphene is compatible with scalable fabrication processes.

The results show that graphene layer number controls the balance between sensitivity and noise. 
Monolayer graphene provides the strongest photogating modulation and therefore the highest sensitivity, whereas few-layer graphene partly screens the substrate field but can improve electrical stability and reduce noise. 
Thus, number of layers should be considered as an optimization parameter.

The detectivity in the present work was not intentially optimized: 
The devices were measured at an ambient conditions, without encapsulation,mpedance optimisation by lateral geometry, interface engineering, additional absorbing layers, or dedicated noise-reduction strategies, increase current through the graphene channel in order to maximize the photoresponse. 
Therefore, the obtained detectivity values of the order of \(10^8\) Jones should be considered as a baseline for simple CVD graphene/SiO$_2$/Si photogating structures rather than as a performance limit. 

Several approaches may improve the detectivity of the CVD-graphene based photogating detectors. 
First, controlled cleaning, annealing, passivation, encapsulation, and operation in inert atmosphere can improve the stability, reproducibility, and detectivity of the devices~\cite{Suk2013polymer,Suhail2017residue,Tyagi2022ultraclean}.

Second, the capacitive coupling between the Si substrate and graphene can be increased. 
A thinner SiO$_2$ layer or a dielectric with higher $\epsilon$ would allow the same photoinduced charge in the substrate to produce a stronger modulation of the graphene resistance. 
This approach should increase the sensitivity of photogating-based devices, but it must be combined with leakage-current suppression.

Third, additional absorbing or charge-trapping layers can be introduced near graphene to enhance the photogating gain. 
This strategy has been successfully used in graphene/MoS$_2$, graphene--PbS quantum-dot, and hBN-assisted graphene photodetectors~\cite{defazio2016high,zheng2023high,fukushima2022graphene}.

Finally, the device geometry can be optimized to improve carrier collection and reduce parasitic contributions. 
Graphene/silicon photodetectors with engineered geometries have shown that device design can strongly affect responsivity~\cite{riazimehr2019high,liu2023high}. 
Therefore, further optimization should balance the photoresponse amplitude, gate coupling, leakage current, and noise level.

\section{Conclusions}

In this work, photoconductive response of mono-, bi-, and trilayer CVD graphene devices on Si/SiO$_2$ substrates was investigated. 
 Devices in which graphene was transferred onto pre-patterned metal contacts demonstrated a stronger photoresponse and were used for further spectral measurements.  Monolayer graphene provided the highest sensitivity, while trilayer graphene showed a lower photoresponse amplitude but a relatively smaller decrease in detectivity.

The origin of the photoresponse is the substrate-induced photogating, as was proven by the observation of a photoresponse in the 405--808~nm spectral range, alongside with the absence of a detectable response at 1550~nm, the strongly sublinear dependence on optical power density, the correlation between the graphene photoresponse and the graphene--Si charging current, and the gate-voltage dependence of the photoresponse. Our results demonstrate that scalable CVD graphene on semiconductor/insulator substrates can be used as a platform for sensitive photogating-based photodetectors.



\begin{acknowledgments}
Microfabrication and optoelectronic measurements were performed at the P.N. Lebedev Physical institute Shared Facility Center and were supported by the Russian Science Foundation under Grant No.~25-42-01036. The transfer process and optimization of obtaining 2L and 3L graphene structures were performed using the scientific equipment of the Center for Collective Use of the National Research Nuclear University MEPhI "Heterostructural Microwave Electronics and Physics of Wide-band Semiconductors"  and supported by the Ministry of Science and Higher Education of the Russian Federation within the framework of the FSWU-2024-0014 project. The authors are grateful to Vadim S. Khrapai and Alexander S. Melnikov for suggesting the direct measurement of the photorecharge current. The authors also thank Anton V. Ikonnikov for valuable discussions,  RusGraphene LLC, particulary Arslan A. Galiullin and Maxim G. Rybin for providing the graphene synthesis setup used in this work.  

\end{acknowledgments}

\bibliography{aipsamp}

@PREAMBLE{
 "\providecommand{\noopsort}[1]{}" 
 # "\providecommand{\singleletter}[1]{#1}%" 
}

@article{li2009large,
  title={Large-area synthesis of high-quality and uniform graphene films on copper foils},
  author={Li, Xuesong and Cai, Weiwei and An, Jinho and Kim, Seyoung and Nah, Junghyo and Yang, Dongxing and Piner, Richard and Velamakanni, Aruna and Jung, Inhwa and Tutuc, Emanuel and Banerjee, Sanjay K. and Colombo, Luigi and Ruoff, Rodney S.},
  journal={Science},
  volume={324},
  number={5932},
  pages={1312--1314},
  year={2009},
  publisher={American Association for the Advancement of Science}
}

@article{reina2009large,
  title={Large area, few-layer graphene films on arbitrary substrates by chemical vapor deposition},
  author={Reina, Alfonso and Jia, Xiaoting and Ho, John and Nezich, Daniel and Son, Hyungbin and Bulovic, Vladimir and Dresselhaus, Mildred S. and Kong, Jing},
  journal={Nano Letters},
  volume={9},
  number={1},
  pages={30--35},
  year={2009},
  publisher={American Chemical Society}
}

@article{gomezdearco2010continuous,
  title={Continuous, highly flexible, and transparent graphene films by chemical vapor deposition for organic photovoltaics},
  author={Gomez De Arco, Lewis and Zhang, Yi and Schlenker, Cody W. and Ryu, Koungmin and Thompson, Mark E. and Zhou, Chongwu},
  journal={ACS Nano},
  volume={4},
  number={5},
  pages={2865--2873},
  year={2010},
  publisher={American Chemical Society}
}

@article{wu2011high,
  title={High-frequency, scaled graphene transistors on diamond-like carbon},
  author={Wu, Yanqing and Lin, Yu-Ming and Bol, Ageeth A. and Jenkins, Keith A. and Xia, Fengnian and Farmer, Damon B. and Zhu, Yu and Avouris, Phaedon},
  journal={Nature},
  volume={472},
  number={7341},
  pages={74--78},
  year={2011},
  publisher={Nature Publishing Group}
}

@article{defazio2016high,
  title={High Responsivity, Large-Area Graphene/MoS$_2$ Flexible Photodetectors},
  author={De Fazio, Domenico and Goykhman, Ilya and Yoon, Duhee and Bruna, Matteo and Eiden, Anna and Milana, Silvia and Sassi, Ugo and Barbone, Matteo and Dumcenco, Dumitru and Marinov, Kolyo and Kis, Andras and Ferrari, Andrea C.},
  journal={ACS Nano},
  volume={10},
  number={9},
  pages={8252--8262},
  year={2016},
  publisher={American Chemical Society},
  doi={10.1021/acsnano.6b05109}
}

@article{nair2008fine,
  title={Fine structure constant defines visual transparency of graphene},
  author={Nair, Rahul R. and Blake, Peter and Grigorenko, Alexander N. and Novoselov, Kostya S. and Booth, Timothy J. and Stauber, Tobias and Peres, Nuno M. R. and Geim, Andre K.},
  journal={Science},
  volume={320},
  number={5881},
  pages={1308},
  year={2008},
  publisher={American Association for the Advancement of Science}
}

@article{kuzmenko2008universal,
  title={Universal optical conductance of graphite},
  author={Kuzmenko, A. B. and van Heumen, Erik and Carbone, Fabrizio and van der Marel, Dirk},
  journal={Physical Review Letters},
  volume={100},
  number={11},
  pages={117401},
  year={2008},
  publisher={American Physical Society}
}

@article{mak2008measurement,
  title={Measurement of the optical conductivity of graphene},
  author={Mak, Kin Fai and Sfeir, Matthew Y. and Wu, Yang and Lui, Chun Hung and Misewich, James A. and Heinz, Tony F.},
  journal={Physical Review Letters},
  volume={101},
  number={19},
  pages={196405},
  year={2008},
  publisher={American Physical Society}
}

@article{bonaccorso2010graphene,
  title={Graphene photonics and optoelectronics},
  author={Bonaccorso, Francesco and Sun, Zhipei and Hasan, Tawfique and Ferrari, Andrea C.},
  journal={Nature Photonics},
  volume={4},
  number={9},
  pages={611--622},
  year={2010},
  publisher={Nature Publishing Group}
}

@article{xia2014graphene,
  title={Graphene and graphene-like two-dimensional materials in photodetection: mechanisms and methodology},
  author={Xia, Fengnian and Mueller, Thomas and Lin, Yu-ming and Valdes-Garcia, Alberto and Avouris, Phaedon},
  journal={Nature Nanotechnology},
  volume={4},
  number={12},
  pages={839--843},
  year={2009},
  publisher={Nature Publishing Group}
}

@article{koppens2014photodetectors,
  title={Photodetectors based on graphene, other two-dimensional materials and hybrid systems},
  author={Koppens, Frank H. L. and Mueller, Thomas and Avouris, Phaedon and Ferrari, Andrea C. and Vitiello, Miriam S. and Polini, Marco},
  journal={Nature Nanotechnology},
  volume={9},
  number={10},
  pages={780--793},
  year={2014},
  publisher={Nature Publishing Group}
}

@article{freitag2013photoconductivity,
  title={Photoconductivity of biased graphene},
  author={Freitag, Marcus and Low, Tony and Xia, Fengnian and Avouris, Phaedon},
  journal={Nature Photonics},
  volume={7},
  number={1},
  pages={53--59},
  year={2013},
  publisher={Nature Publishing Group}
}

@article{guo2016high,
  title={High-performance graphene photodetector using interfacial gating},
  author={Guo, Xitao and Wang, Wenhui and Nan, Haiyan and Yu, Yuanfang and Jiang, Jie and Zhao, Weiwei and Li, Jinhuan and Zafar, Zainab and Xiang, Nan and Ni, Zhonghua and Hu, Weida and You, Yumeng and Ni, Zhenhua},
  journal={Optica},
  volume={3},
  number={10},
  pages={1066--1070},
  year={2016},
  publisher={Optica Publishing Group}
}

@article{jago2019microscopic,
  title={Microscopic origin of the bolometric effect in graphene},
  author={Jago, Roland and Malic, Ermin and Wendler, Florian},
  journal={Physical Review B},
  volume={99},
  number={3},
  pages={035419},
  year={2019},
  publisher={American Physical Society}
}

@article{echtermeyer2014photothermoelectric,
  title={Photo-thermoelectric and photoelectric contributions to light detection in metal-graphene-metal photodetectors},
  author={Echtermeyer, Tim J. and Nene, Prathamesh S. and Trushin, Maxim and Gorbachev, Roman V. and Eiden, Alexander L. and Milana, Silvia and Sun, Zhipei and Schliemann, John and Lidorikis, Elefterios and Novoselov, Kostya S. and Ferrari, Andrea C.},
  journal={Nano Letters},
  volume={14},
  number={7},
  pages={3733--3742},
  year={2014},
  publisher={American Chemical Society}
}

@article{bandurin2018dual,
  title={Dual origin of room temperature sub-terahertz photoresponse in graphene field effect transistors},
  author={Bandurin, Denis A. and Gayduchenko, Igor and Cao, Yuan and Moskotin, Maxim and Principi, Alessandro and Grigorieva, Irina V. and Goltsman, Gregory and Fedorov, Georgy and Svintsov, Dmitry},
  journal={Applied Physics Letters},
  volume={112},
  number={14},
  pages={141101},
  year={2018},
  publisher={AIP Publishing}
}

@article{Suk2013polymer,
  title={Enhancement of the electrical properties of graphene grown by chemical vapor deposition via controlling the effects of polymer residue},
  author={Suk, Ji Won and Lee, Wi Hyoung and Lee, Jongho and Chou, Harry and Piner, Richard D. and Hao, Yufeng and Akinwande, Deji and Ruoff, Rodney S.},
  journal={Nano Letters},
  volume={13},
  number={4},
  pages={1462--1467},
  year={2013},
  publisher={American Chemical Society}
}

@article{Abhilash2015transfer,
  title={Transfer printing of CVD graphene FETs on patterned substrates},
  author={Abhilash, T. S. and De Alba, Roberto and Zhelev, Nikolay and Craighead, Harold G. and Parpia, Jeevak M.},
  journal={Nanoscale},
  volume={7},
  number={33},
  pages={14109--14113},
  year={2015},
  publisher={Royal Society of Chemistry}
}

@article{Tyagi2022ultraclean,
  title={Ultra-clean high-mobility graphene on technologically relevant substrates},
  author={Tyagi, Ayush and Mi\v{s}eikis, Vaidotas and Martini, Leonardo and Forti, Stiven and Mishra, Neeraj and Gebeyehu, Zewdu M. and Giambra, Marco A. and Zribi, Jihene and Fr{\'e}gnaux, Mathieu and Aureau, Damien and Romagnoli, Marco and Beltram, Fabio and Coletti, Camilla},
  journal={Nanoscale},
  volume={14},
  number={6},
  pages={2167--2176},
  year={2022},
  publisher={Royal Society of Chemistry}
}

@article{Kronik1999surface,
  title={Surface photovoltage phenomena: theory, experiment, and applications},
  author={Kronik, Leeor and Shapira, Yoram},
  journal={Surface Science Reports},
  volume={37},
  number={1},
  pages={1--206},
  year={1999},
  publisher={Elsevier}
}

@article{Novikov2010experimental,
  title={Experimental measurement of work function in doped silicon surfaces},
  author={Novikov, A.},
  journal={Solid-State Electronics},
  volume={54},
  number={1},
  pages={8--13},
  year={2010},
  publisher={Elsevier}
}

@article{Konstantatos2012hybrid,
  title={Hybrid graphene--quantum dot phototransistors with ultrahigh gain},
  author={Konstantatos, Gerasimos and Badioli, Michele and Gaudreau, Louis and Osmond, Johann and Bernechea, Maria and de Arquer, F. Pelayo Garcia and Gatti, Fabrice and Koppens, Frank H. L.},
  journal={Nature Nanotechnology},
  volume={7},
  number={6},
  pages={363--368},
  year={2012},
  publisher={Nature Publishing Group}
}

@article{riazimehr2019high,
  title={High responsivity and quantum efficiency of graphene/silicon photodiodes achieved by interdigitating Schottky and gated regions},
  author={Riazimehr, Sarah and Kataria, Satender and Gonzalez-Medina, Jose M. and Wagner, Stefan and Shaygan, Mehrdad and Suckow, Stephan and Ruiz, Francisco G. and Engstrom, Olof and Godoy, Andres and Lemme, Max C.},
  journal={ACS Photonics},
  volume={6},
  number={1},
  pages={107--115},
  year={2019},
  publisher={American Chemical Society}
}

@article{wang2021interfacial,
  title={Interfacial photogating effect for hybrid graphene-based photodetectors},
  author={Wang, Yifei and Ho, Vinh X. and Pradhan, Prashant and Cooney, Michael P. and Vinh, Nguyen Q.},
  journal={ACS Applied Nano Materials},
  volume={4},
  number={8},
  pages={8539--8545},
  year={2021},
  publisher={American Chemical Society}
}

@article{fukushima2022graphene,
  title={Graphene-based deep-ultraviolet photodetectors with ultrahigh responsivity using chemical vapor deposition of hexagonal boron nitride to achieve photogating},
  author={Fukushima, Shoichiro and Fukamachi, Satoru and Shimatani, Masaaki and Kawahara, Kenji and Ago, Hiroki and Ogawa, Shinpei},
  journal={Optical Materials Express},
  volume={12},
  number={5},
  pages={2090--2101},
  year={2022},
  publisher={Optica Publishing Group}
}

@article{zheng2023high,
  title={High-performance graphene-PbS quantum dots hybrid photodetector with broadband response and long-time stability},
  author={Zheng, Jiajin and Di, Wanchao and Bao, Beibei and Lu, Jiaqi and Yu, Kehan and Wei, Wei},
  journal={Applied Physics B},
  volume={129},
  number={3},
  pages={43},
  year={2023},
  publisher={Springer}
}

@article{liu2023high,
  title={High-performance broadband graphene/silicon/graphene photodetectors: From x-ray to near-infrared},
  author={Liu, Xinyu and Ning, Hao and Lv, Jianhang and Liu, Lixiang and Peng, Li and Tian, Feng and Bodepudi, Srikrishna Chanakya and Wang, Xiaochen and Cao, Xiaoxue and Dong, Yunfan and Fang, Wenzhang and Wu, Shaoxiong and Hu, Huan and Yu, Bin and Xu, Yang},
  journal={Applied Physics Letters},
  volume={122},
  number={7},
  pages={071105},
  year={2023},
  publisher={AIP Publishing}
}

@article{liu2014graphene,
  title={Graphene photodetectors with ultra-broadband and high responsivity at room temperature},
  author={Liu, Chang-Hua and Chang, You-Chia and Norris, Theodore B. and Zhong, Zhaohui},
  journal={Nature Nanotechnology},
  volume={9},
  number={4},
  pages={273--278},
  year={2014},
  publisher={Nature Publishing Group}
}

@article{zhang2014ultrahigh,
  title={Ultrahigh-gain photodetectors based on atomically thin graphene-MoS$_2$ heterostructures},
  author={Zhang, Wenjing and Chuu, Chih-Piao and Huang, Jing-Kai and Chen, Chang-Hsiao and Tsai, Meng-Lin and Chang, Yung-Huang and Liang, Chi-Te and Chen, Yu-Ze and Chueh, Yu-Lun and He, Jr-Hau and Chou, Mei-Yin and Li, Lain-Jong},
  journal={Scientific Reports},
  volume={4},
  pages={3826},
  year={2014},
  publisher={Nature Publishing Group}
}

@article{ievleva2025metastable,
  title={Metastable States of 2D-Material-on-Metal-Islands Structures Revealed by Thermal Cycling},
  author={Ievleva, Valeriya A. and Prudkoglyad, Valery A. and Morgun, Leonid A. and Kuntsevich, Aleksandr Yu.},
  journal={Micromachines},
  volume={16},
  number={12},
  pages={1385},
  year={2025},
  publisher={MDPI},
  doi={10.3390/mi16121385}
}

@article{cha2022bottom,
  title={A Bottom-Electrode Contact: The Most Suitable Structure for Graphene Electronics},
  author={Cha, Jongin and Son, Jangyup and Hong, Jongill},
  journal={Advanced Materials Interfaces},
  volume={9},
  number={6},
  pages={2102207},
  year={2022},
  publisher={Wiley},
  doi={10.1002/admi.202102207}
}

@article{teneh2012,
  title={Spin-Droplet State of an Interacting 2D Electron System},
  author={Teneh, N. and Kuntsevich, A. Yu. and Pudalov, V. M. and Reznikov, M.},
  journal={Physical Review Letters},
  volume={109},
  number={22},
  pages={226403},
  year={2012},
  publisher={American Physical Society},
  doi={10.1103/PhysRevLett.109.226403}
}

@article{kuntsevich2015,
  title={Strongly correlated two-dimensional plasma explored from entropy measurements},
  author={Kuntsevich, A. Yu. and Tupikov, Y. V. and Pudalov, V. M. and Burmistrov, I. S.},
  journal={Nature Communications},
  volume={6},
  pages={7298},
  year={2015},
  publisher={Nature Publishing Group},
  doi={10.1038/ncomms8298}
}

@article{Suhail2017residue,
  title={Reduction of Polymer Residue on Wet-Transferred CVD Graphene Surface by Deep UV Exposure},
  author={Suhail, Ahmed and Islam, Kamrul and Li, Bobo and Jenkins, David and Pan, Guobing},
  journal={Applied Physics Letters},
  volume={110},
  number={18},
  pages={183103},
  year={2017},
  publisher={AIP Publishing},
  doi={10.1063/1.4983185}
}

@article{brattain1956combined,
  title={Combined Measurements of Field Effect, Surface Photo-Voltage and Photoconductivity},
  author={Brattain, W. H. and Garrett, C. G. B.},
  journal={The Bell System Technical Journal},
  volume={35},
  pages={1019--1040},
  year={1956},
  doi={10.1002/j.1538-7305.1956.tb03816.x}
}

@article{johnson1957measurement,
  title={Measurement of Minority Carrier Lifetimes with the Surface Photovoltage},
  author={Johnson, E. O.},
  journal={Journal of Applied Physics},
  volume={28},
  number={11},
  pages={1349--1353},
  year={1957},
  doi={10.1063/1.1722650}
}

@article{quillet1960surface,
  title={L'effet photovoltaique de surface dans le silicium et son application {\`a} la mesure de la dur{\'e}e de vie des porteurs minoritaires},
  author={Quillet, A. and Gorsar, P.},
  journal={Journal de Physique et le Radium},
  volume={21},
  pages={575--580},
  year={1960},
  doi={10.1051/jphysrad:01960002107057500}
}

@article{goodman1961method,
  title={A Method for the Measurement of Short Minority Carrier Diffusion Lengths in Semiconductors},
  author={Goodman, A. M.},
  journal={Journal of Applied Physics},
  volume={32},
  number={12},
  pages={2550--2552},
  year={1961},
  doi={10.1063/1.1728351}
}

@article{frankl1967theory,
  title={Theory of the Small-Signal Photovoltage at Semiconductor Surfaces},
  author={Frankl, D. R. and Ulmer, E. A.},
  journal={Surface Science},
  volume={6},
  number={1},
  pages={115--123},
  year={1967},
  doi={10.1016/0039-6028(67)90017-9}
}

@article{lam1971surfaceI,
  title={Surface-State Density and Surface Potential in MIS Capacitors by Surface Photovoltage Measurements. I},
  author={Lam, Y. W.},
  journal={Journal of Physics D: Applied Physics},
  volume={4},
  number={9},
  pages={1370--1375},
  year={1971},
  doi={10.1088/0022-3727/4/9/318}
}

@article{lam1971surfaceII,
  title={Surface-State Density and Surface Potential in MIS Capacitors by Surface Photovoltage Measurements. II},
  author={Lam, Y. W. and Rhoderick, E. H.},
  journal={Journal of Physics D: Applied Physics},
  volume={4},
  number={9},
  pages={1376--1389},
  year={1971},
  doi={10.1088/0022-3727/4/9/319}
}

@article{borisenko2025raman,
  title={Raman Spectroscopy of Multilayer Graphene Structures with Various Twist Angles Between Layers},
  author={Borisenko, D. P. and Kargin, N. I. and Rybin, M. G.},
  journal={Journal of Applied Spectroscopy},
  volume={92},
  number={4},
  pages={762--769},
  year={2025},
  publisher={Springer},
  doi={10.1007/s10812-025-01970-6}
}

@article{rybin2024rapid,
  title={Rapid synthesis of CVD graphene with controllable charge carrier mobility},
  author={Rybin, Maxim G. and Guberna, Evgeniy A. and Obraztsova, Ekaterina A. and Kondrashov, Ivan and Kurkina, Irina I. and Smagulova, Svetlana A. and Obraztsova, Elena D.},
  journal={Carbon Trends},
  volume={15},
  pages={100349},
  year={2024},
  publisher={Elsevier},
  doi={10.1016/j.cartre.2024.100349}
}

\end{document}